# Water harvesting from Soils by Solar-to-Heat Induced Evaporation and Capillary Water Migration

Xiaotian Li[†], Guang Zhang[†], Chao Wang, Lichen He, Yantong Xu, Rong Ma and Wei Yao*

*Department of Space sciences, Qian Xuesen Laboratory of Space Technology, China Academy of Space Technology, Beijing 100094, China.*

[†]*These authors contributed equally to this work.*

*Correspondence and requests for materials should be addressed to W. Y (yaowei@qxslab.cn)

**ABSTRACT**

Fresh water scarcity is one of the critical challenges for global sustainable development. Several novel water resources such as passive seawater solar desalination and atmospheric water harvesting have made some progress in recent years. However, no investigation has referred to harvesting water from shallow subsurface soils, which are potential huge water reservoirs. Here, we introduce a method of solar-driven water harvesting from soils, which can provide cheap fresh water in impoverished, arid and decentralized areas. The concentrated solar energy is used to heat the soils to evaporate the soil moisture. Then vapors flow to the condenser through tubes and condense as freshwater. Sustainable water harvesting is realized by water migration due to capillary pumping effect within soils. In the laboratory condition, an experimental setup is designed and its water-harvesting ability from soils is investigated. The maximum water mass harvesting rate was 99.8 g $h^{-1}$. In about 12 h, the total harvesting water could be as high as about 900 ml. The whole process is solar-driven and spontaneous without other mechanical or electrical ancillaries. The water harvesting rate under one sun energy flux (1 kW $m^{-2}$) is estimated to be about 360 g $h^{-1}$ with a 1 $m^2$ solar concentrator. Our proposal provides a potential onsite and sustainable fresh water supply solution to deal with the water scarcity problem.

## 1. INTRODUCTION

Water is essential in our lives, whereas the scarcity of fresh water is still an increasingly devastating global problem (Everard, 2019). Ensuring access to clean water for all mankind is one of the 17 sustainable development goals, which is essential and challenging("Goal 6: Ensure access to water and sanitation for all," n.d.). The projected water shortfall for 2030 is almost 2000 billion m$^3$, more than 20% of the projected global needs(Chiavazzo et al., 2018a; Grum et al., 2016; Kallenberger and Fröba, 2018; Tu et al., 2018a). 1.1 billion people still have limited access to clean fresh water according to a report by WHO and predicted to be worsen with the increasing demands due to population growth and economic development("1 in 3 people globally do not have access to safe drinking water – UNICEF, WHO," n.d.). In particular, the fresh water shortage is more severe in arid, impoverished, remote areas, and at emergency times. Nowadays, seawater desalination and atmospheric water harvesting technologies are regarded as two critical ways to alleviate globe water shortage.

In specific, mature desalination technologies such as reverse osmosis and bulk desalination are fossil energy cost and environment impact. In addition, large capital consumption and cumbersome infrastructure make it unavailable in remote rural areas. Recently, increasing interest has been directed toward high efficiency, low-cost, scalable, green and renewable fresh water desalination, among them solar-driven interfacial desalination(Bae et al., 2015; Cao et al., 2019; Ghasemi et al., 2014; G. Li et al., 2018; Siva Reddy et al., 2013; Tao et al., 2018; Wang et al., 2019; Zhang et al., 2015) and solar-driven passive membrane distillation (Chiavazzo et al., 2018b)are especially attractive. But after desalination, high-salinity brines are hard to deal with, which may cause ecological problems. Moreover, desalination is infeasible in the landlocked areas without seawater

resources(Miller et al., 2014), which constrains its application. The other way is to harvest water from the atmosphere. Typical AWGs (atmospheric water generations) such as fog water collection(Al-hassan, 2009; Fessehaye et al., 2014; Marzol Jaén, 2002) and dew water collection(Seo et al., 2014) is hard to harvest water from low humidity air, while novel adsorption-based AWGs employing inorganic salt(Ji et al., 2007; Kallenberger and Fröba, 2018; R. Li et al., 2018), zeolite, brines and MOFs (metal–organic frameworks) (Fathieh et al., 2018; Kalmutzki et al., 2018; Kim et al., 2018; LaPotin et al., 2019; Rieth et al., 2017; Umans et al., 2017)have recently flourished and shows promising prospects even in arid atmosphere. However, the thermodynamic performance of adsorption-based AWGs is relatively low and some absorbers such as MOFs are still unstable and unhealthy(Tu et al., 2018b).

Apart from seawater and atmospheric vapor, water storage in shallow subsurface soil between 0 - 2 m is tremendous. It is reported that average soil water content (SWC) between 0 - 2 m underground is normally in the range of 1.25 % - 25 % gravimetric content (or approximately 2% - 40% volumetric content) in most areas("https://www.geo.tuwien.ac.at/insitu/data_viewer/#," n.d.; Liu and Xie, 2013; Qin et al., 2013). Utilizing this huge water resources to will benefit a lot to alleviate water shortage. However, no investigations have been made to extracting water from the shallow subsurface soil to date. In this work, we demonstrated that fresh water extraction from soil is feasible for the first time. The integrated experimental setup could effectively harvest fresh water from soils with different SWC. The maximum water mass harvesting rate was 0.1 kg h$^{-1}$. In about 12 h, the total harvesting water could be as high as about 0.9 kg, which could approximately meet a single person's drinking water needs. Then the maximum water harvesting rate of 20 wt.% SWC soil was estimated to be about 0.36 kg h$^{-1}$ under one sun (1 kW m$^{-2}$) with a 1 m$^2$ solar concentrator.

Our experimental and theoretical results also demonstrated that the soil moisture could be pumped to the evaporation interface by capillary force, which enabled quick water replenishment for continuous water evaporation. Herein, our work opened a new avenue to augment fresh water supply to alleviate the water shortage with totally solar energy.

## 2. METHODS AND MATERIALS

### 2.1 The scheme for solar water mining from soils

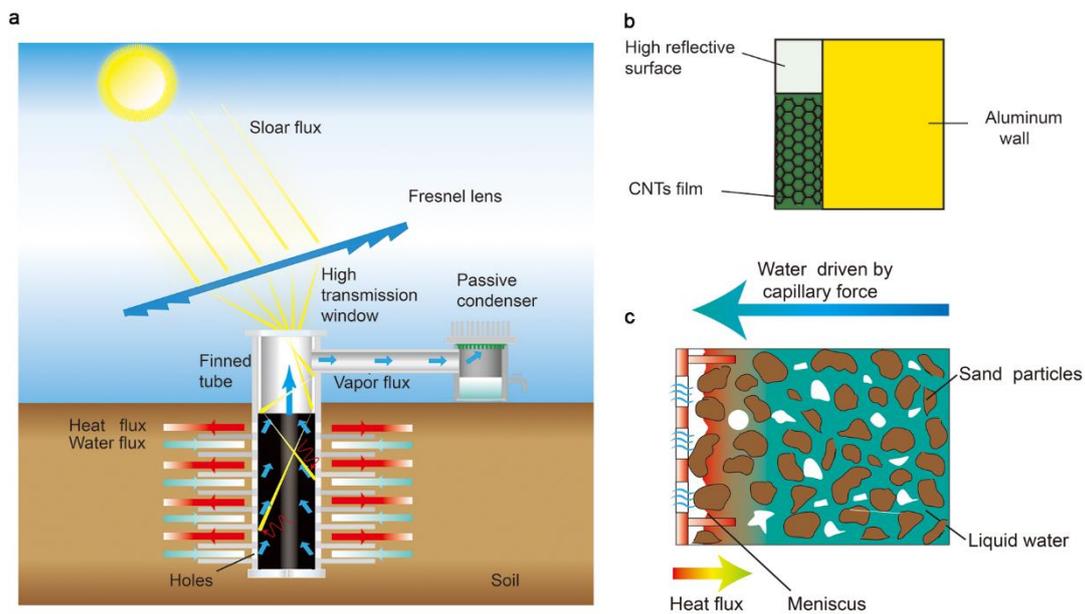

**Fig. 1 The scheme for harvesting water from soil. a,** The setup of our proposal consists of a solar concentrator, a metallic hollow finned tube, and a condensation collector. **b,** Schematic diagrams of the inside-walls of the finned tube, where the upper part is covered by the high reflective surface and the lower side is coated by CNT films. **c,** Schematic diagrams of the as-assembled embedded hollow finned tube and capillary water migration.

The setup of our proposal consists of a solar concentrator, a metallic hollow finned tube, and a condensation collector. As depicted in **Fig.1a**, a finned metallic tube is buried vertically into the soils. The solar radiation is focused by a Fresnel lens into the tube from the top transparent window and is reflected from the upper part to the lower part inside the tube. Then, the solar irradiation is approximately completely absorbed by a highly absorbable coating. Fins enhance the heat transfers to the soils and evaporate the soil moisture. Intensive small holes on the wall allow vapors to flow into the tube, meanwhile, the soil moisture will migrate from the far side to the heated tube due to capillary force. Once vapors move up to the passive condenser through a pipe, they are condensed to fresh water.

As shown in **Fig. 1b,** the upper side is polished to form a high solar reflective surface. Then the lower side is coated by multi-walled carbon nanotube (CNT) films for perfect adsorption of solar radiation(Gokhale et al., 2014). The fabrication, characterization, and properties of the CNT films are provided in the **Supplementary Material**. The process for heat and moisture transfer during solar mining is depicted in **Fig. 1c**. The heat continues transferring to the soils from the finned tube. The liquid water at the meniscus of water-air interface near the tube continues evaporating to vapors with the tube temperature increases. At the same time, the far side soil moisture transports toward the tube due to the effective capillary force through the tiny porous pathways, which is in analogy with the heat pipe effect(Udell, 1985). Then, the vapor pressure gradient is built between the soil pores and the inside of the finned tube, promoting vapor flux into the finned tube through the small holes on the finned tube wall. Therefore, the accumulated vapors inside the tube provide a positive pressure to drive the net vapor flow to the condenser.

**2.2 Laboratory experiment on harvesting water from sand soils**

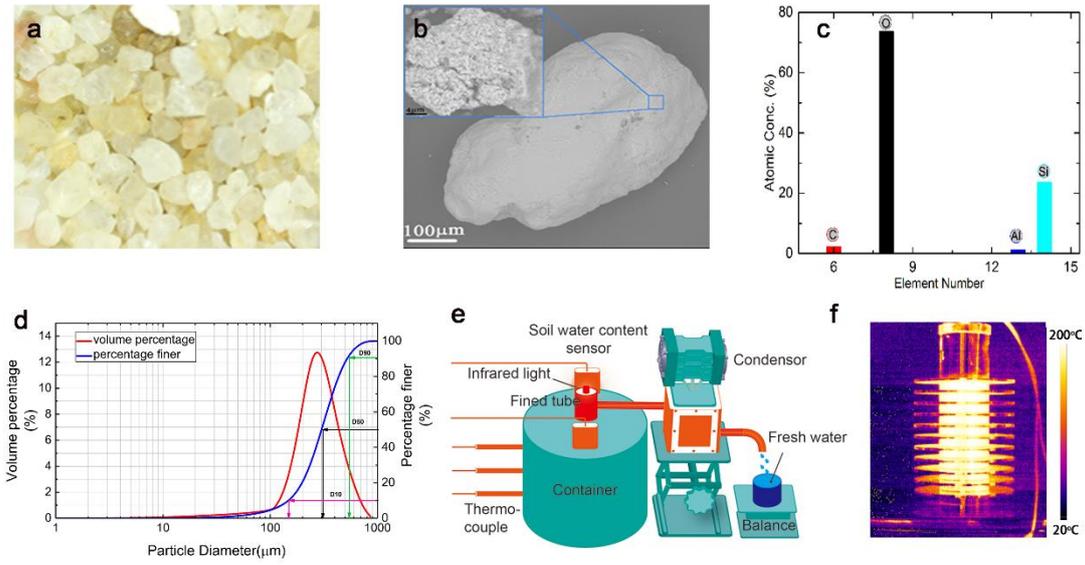

**Fig. 2 Experiment materials and setup. a**, Microscopy picture of porous sand soils. **b**, Scanning electron microscope (SEM) images of a soil particle. Inset: high-magnification SEM image of the soil particle. **c**, EDS characterization of the soil sample, which indicates that the soil mainly consists of Si and O elements; **d**, Particle diameter distribution of soil sample. **e**, The as-assembled laboratory experiment setup for harvesting water from soil. **f**, The IR image of the infrared light heated finned tube, which clearly indicates that the heat mostly concentrated in the parts buried into the soil.

In order to confirm the feasibility of our proposal, an experimental setup was designed and its water-harvesting ability from sand soils was thoroughly investigated. The used sand soils were collected from the Yellow Sea beach (Rizhao, Shandong). These sand soil samples were first treated in an oven at 105 °C for 24 hours and then filtered by a sieve. Some sand soils were put in a plexiglass container. The microscopy picture of the used sand soils shows the porous features of the sand soils (**Fig. 2a**). The scanning electron microscope (SEM) images of a sand particle is shown in **Fig. 2b**. The morphology is irregular and the surface have high level roughness. Then the information of the chemical components is detected by the Energy Dispersive Spectrometer (EDS). The tested results are shown in **Fig. 2c**, which demonstrates that the main composition is $SiO_2$. The particle diameter distribution is shown in **Fig. 2d**. It is shown that particles with a particle size smaller than 512.1 μm is account for 90 % (D90). Then the D50 and D10 and are 300.5 μm and 183.1 μm, respectively.

The as-assembled laboratory experiment setup is shown in **Fig. 2e**. A hollow tube was buried vertically in the sand soils. The hollow tube was made up of aluminum (200 mm in length, 30 mm in diameter and 2 mm wall thickness). The photograph of the hollow tube is shown in Supplementary Information (**Fig. S2**). Intensive holes with 1 mm diameter were cut through the wall. To enhance heat transfer, eleven fins with 10 mm pitch were integrated on the outer walls of the drill, each was 3 mm thick and 30 mm larger than the outer diameter of the drill in the radial direction. The plexiglass container was 250 mm in diameter and 200 mm in length.

An infrared light was used to simulate solar concentrator effect in the laboratory condition. The heating power of the infrared light was controlled by a DC source. The infrared light was 12 mm in diameter and 250 mm in length, of which the light wavelength was 7-14 μm. The power of the infrared heater ranged from 0 - 800 W and could be modified by changing the voltage and current when connected to an adjustable DC power supply. The tests were performed in the laboratory environment where the relative humanity (ΔRH) was 71% and room temperature was 21-24 °C. K-type thermocouples (range from -40 ~ 900 ºC, resolution 0.1 ºC) were used to measure the temperature on the outer wall of the finned tube. They could be divided into 3 groups at 60 mm depth increments from 50 mm below the top of the finned tube. Each group consisted of a pair of thermocouples placed at the same height level but different circumference location in order to minimize the random error. A multiple-channel datalogger was used to acquire the temperature data.

As shown in **Fig. 2f**, the upper wall of the finned tube was less heated and the lower side had a higher temperature, which demonstrated that the heat was mainly absorbed by the lower part. Then the heated vapor reached into an upper closure of condenser, which was mounted with an aluminum fins and sprayed by hydrophobic coatings to enhance the condensation of water vapor. Because the fins enlarged the cold surface for water molecules to collide and the hydrophobic film allows the water droplet to drop rapidly when it grown over the threshold radius. The distilled water accumulated on the damp at bottom and then slipped to the water tank. Changes in water content of soils were measured by two CS655 TDR probes (Campbell scientific tech. co.td) at 10 min intervals for 7 hours. The distribution of the two probes were 10 cm and 12 cm, respectively, from the center of the finned tube. The data collection was implemented by a CR800 datalogger (Campbell scientific tech. co. td). The moisture Migration experiment was conducted under the 20 % moisture condition.

## 3 RESULTS AND DISCUSSION

### 3.1 Water harvesting performance of our setup

Here, in order to identify the water harvesting ability of our proposal, sand soils with different SWC (5%, 10%, and 20%) were prepared and tested for water harvesting experiments. Furthermore, every soil sample was heated by three different infrared light power (200 W, 250 W, and 300 W). Here, the temperature profiles of the finned tube walls were measured. The temperature profiles at 200 W is shown in **Fig. 3a-c**. It can be seen that temperature raised rapidly then a plateau appeared and lasted for several hours. The plateau stayed for several hours regardless of different input powers and different water contents, meaning that the water was boiling steadily by consuming large quantity of latent heat, also indirectly revealed that the moisture migration to these areas took place continuously. After that, the temperature increased rapidly again because the mean water content declined during the process resulting that the vaporizing rate was faster than the liquid water migration rate into the vaporization zone, which lead to a fast temperature rising of the porous matrix. At 10% and 20% water content, the temperature arrived to the boiling point firstly, and then stayed steady for several hour. The steady time was longer when water content was higher. In our study the maximum mean temperature of the finned tube was 350 °C, which is far below the melting temperatures of the aluminum and CNTs.

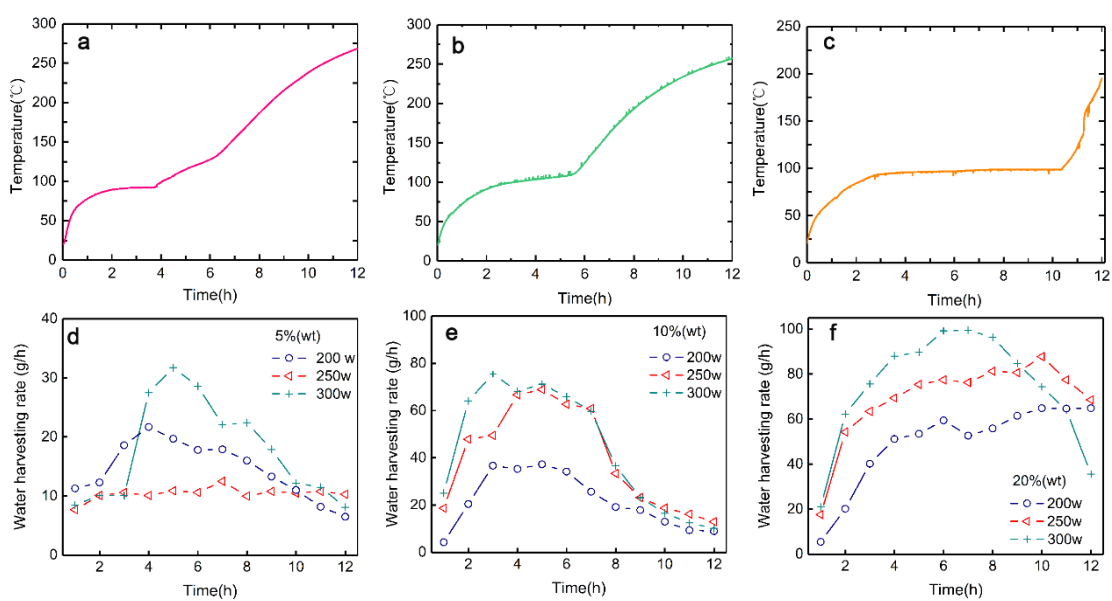

**Fig. 3 Water harvesting performances. a**, Temperature of finned tube walls at 5% water content, 200 W. **b**, Temperature of finned tube walls at 10% water content, 200 W. **c**, temperature of finned tube walls at 20% water content, 200 W. **d**, Water harvesting rate (water harvesting mass per hour, g h$^{-1}$) against different power at 5% water content. **e**, Water harvesting rate against different power at 10% water content. **f**, Water harvesting rate against different power at 20% water content.

The water harvesting performances of our as-assembled setup are shown in **Fig. 3d-f**. As shown in **Fig. 3d**, when the soil is relatively arid (SWC ~ 5%), our setup could effectively extract fresh water from the arid soil at three different heating powers (200 W, 250 W, and 300 W). The water collection mass increased with the heating power. In addition, the water collection rate (water harvesting mass per hour, g h$^{-1}$) firstly increased and then reached to a stable value. In 12 h, the total water collection mass were 210.7 g, 124.9 g, and 174.3 g, respectively. At 10% water content, the water harvesting rate firstly increased to stable value and then sharply decrease at the time of about 7 h (**Fig. 3e**). In 12 h, the total water collection mass were 262.5 g, 479.9 g, and 528.9 g, respectively. This result clearly indicated that the total water collection mass in 12 h increased with heating powers. For 20% water content, the water harvesting rate only had a sharp decrease at the heating power of 300 W (**Fig. 3f**). However, at heating powers of 200 W and 250 W, the water harvesting rate increased to stable values in about 2 h and then kept the stable values for 7 h. In 12 h, the total water collection mass were 592.7 g, 8298 g, and 890.8 g, respectively. Then the water harvesting rate decreased because the container had limited sizes and couldn't keep stable water migration. The total water harvesting masses for different soils with different SWC at different heating powers are summarized in **Fig. 4a**. These results demonstrate that, after the 12 h experiment, almost half of the total water in the soil samples could be extracted by our method.

**Fig. 3d** shows that at 5% soil content, the effect of input power was no significant on harvesting rate maybe because there were few liquid bridges in the samples to replenish the water from the far side to the vaporization zones and the main water migration mechanism was vapor transport. For the soils with initial water content of 10% and 20%, the water harvesting rate positively correlated to input heating powers (**Fig. 3e** and **f**,). A previous work showed that the water in soil pores tend to joint together at higher mean water content and form continuous liquid phase in most regions(Whitaker, n.d.). At the same time, the higher input power could heat up soils faster and increased the vaporization rate. Under the power limits, a larger evaporation rate induced by more intense power, leading to higher water migration flux toward the finned tube when the water was continuous connected. The total water production in this study is expected to increase as water content increases at the same input power, which indicates that higher water content makes easier water harvesting.

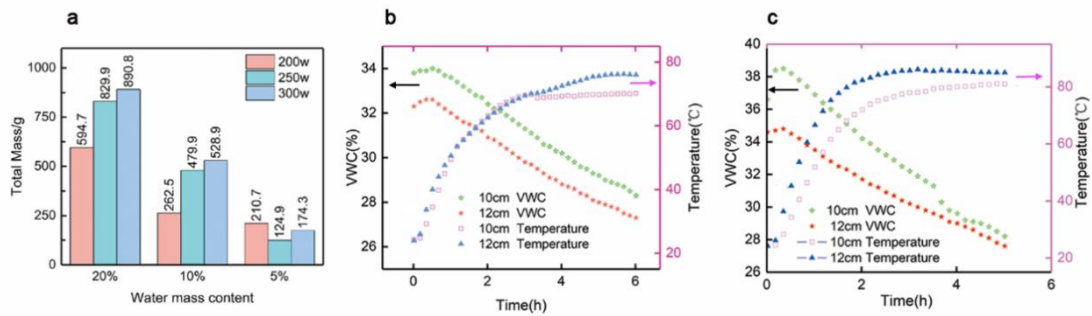

**Fig. 4 Total water harvesting masses and soil moisture migration results. a,** The total harvesting water mass in about 12 h. **b,** Water content change at 10 cm and 12 cm away from finned tube center at 20% water content and 200 W. **c,** Water content change at 10 cm and 12 cm away from finned tube center at 20% water content and 300 W.

**Fig. 4b** and **c** show the water migration experimentsal results of water migration. The gravimetric soil water content was transformed to the volume soil water content (VWC) to show the migration of soil moisture during our experiment. The soil moisture transported away from the tube at first because the water vapors transports to the colder surface as the temperature of the tube increases. However, due to the increase of capillary force resulting from the increasing evaporation rate, this progress only sustained for about several minutes. The soil moisture migration rate caused by capillary force is faster than vapor flux due to temperature gradient. So, the total soil moisture migration is toward the tube, resulting in the decreasing water content of the soil at the far side. The results also show that the migration rate increased as the input power increased. Because in our experiment condition, higher input power induces higher capillary force. Thus, we demonstrated the continuous water migration to the tube during our experiment. We conducted the water migration experiment for 6 hours because the temperature of soil water content sensor couldn't work for long time when the temperature is higher than 70 ℃.

**3.2 Numerical simulation results**

To reveal the water distribution in the sand during experiment, COMSOL Multiphysics software was used to simulate the process at 20% water content and 200 W. The model results were shown in **Fig.5** (a-e). A coupled porous media heat and mass model(Hongbing et al., 2015; Jin et al., 2018; Nassar and Horton, 1997; Wang et al., 2015) was used to describe the temperature and moisture distribution in the sand sample, parameters can be found in the listed references. In order to simplify calculation, a two-dimensional axisymmetric domain was used. (**Supplementary Information Fig. S3)**. The water transport process was modeled. The water content decreased in the vicinity of the finned tube as evaporation was in progress. At the same time, the water content at the far side away from the finned tube also decreased due to water migration induced by capillary force toward the center. Simultaneously, the simulation results of the temperature distribution at 12 h in the sample showed a good agreement with our experiment results. The heat was mainly concentrated around the finned tube because phase change and water migrations kept most heat from transporting deeper into the sample. In another word, the sample was locally heated. The highest temperature of the finned tube was about 260 ℃ and the temperature at the edge of the sample was nearly 50 ℃.

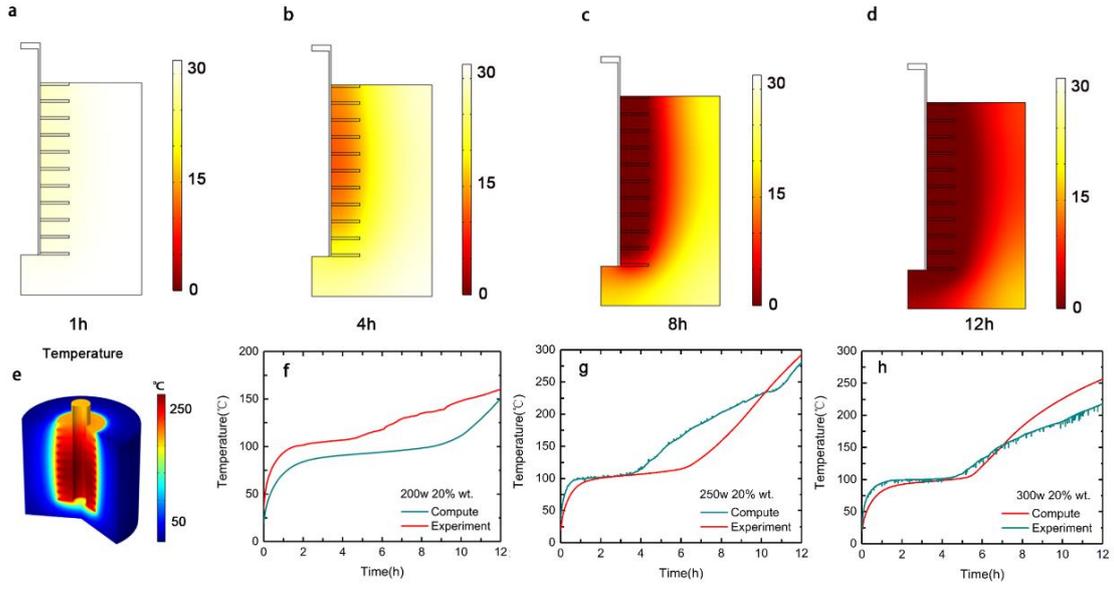

**Fig 5. Numerical results. a,** the moisture content distribution of soils at heating power of 200 W at 1st hour. **b,** moisture content distribution of soils at heating power of 200 W at 4th hour. **c,** moisture content distribution of soils at heating power of 200 W at 8th hour. **d,** moisture content distribution of soils at heating power of 200 W at 12th hour. **e,** temperature distribution of the setup at the 12th hour. **f,** temperature of walls at 200 W and 20% wt. **g,** temperature of walls at 250 W and 20% wt. **h,** temperature of walls at 300 W and 20% wt.

### 3.3 Maximum water harvesting capability analysis

It is assumed that in the process of water abstraction, the liquid water in the sand is continuously distributed, and the evaporation rate is considered to be equal to the capillary flow rate, and the liquid can be continuously replenished, and the temperature of the boiling liquid surface is assumed to be constant, and the influence of temperature on the capillary force is considered to be negligible. And then proceed to the discussion below. In sand, the interfacial capillary force of the pore liquid-non-wetting phase is:

$$p_c = \frac{2\sigma \cos \theta}{r_{eff}} \quad (1)$$

Assume that the sand particles are spherical, then $r_{eff}=0.21d_p$, $d_p$ and is the diameter of the particles(Management, n.d.).

In analog with heat pipes, the capillary force should be equal to the hydraulic pressure drop:

$$p_c = \Delta p_l \tag{2}$$

The mass flow rate of liquid water in sand under the one-dimension axisymmetric condition can be described by:

$$\Delta p_1 = \frac{\dot{m}v}{2\pi K}\ln\frac{r_2}{r_1} \tag{3}$$

Where $\dot{m}$ is the mass flow caused by capillary force, $K$ is the permeability, and $v$ is the kinematic viscosity. $r_1$ is the radius of the finned tube pipe, and $r_2$ is the radius of the soil cylinder.

$K$ can be described by Kozeny–Carman equation:

$$K = d_p^2\varepsilon^3/180(1-\varepsilon)^2 \tag{4}$$

The capillary force is equal to the pressure drop of the liquid flow, and the relationship between the capillary flow and the capillary force can be described by the following formula:

$$\frac{2\sigma\cos\theta}{r_{eff}} = \frac{\dot{m}v}{2\pi K}\ln\frac{r_2}{r_1} \tag{5}$$

Assuming that the heat transferred to the porous substrate is negligible during steady evaporation, the input energy is used to heat the liquid water flowing into the control body and evaporate the water.

$$q'' = \dot{m}_v h_{lg} + \dot{m}c_{p1}(T_v - T_l) \tag{6}$$

Where $\dot{m}_v$ is the steam flow, $h_{lg}$ is the evaporation enthalpy, $c_{p1}$ is the specific heat capacity, and $T_v$ is the boiling point and $T_l = 20$ ℃.

As the evaporation progress proceeds, the curvature of the meniscus gradually increases, and the contact angle gradually decreases to zero. At this time, the capillary force will reach a maximum value. When the evaporation rate and capillary flow are equal, the relationship between maximum capillary force and maximum heat flux is obtained.

$$q_{max}'' = \frac{4\sigma\pi K}{\nu r_{eff}\ln\frac{r_2}{r_1}}[h_{lg} + c_{pl}(T_v - T_l)] \qquad (7)$$

Substituting $p_{cmax}$ into the equation (7) gives the best energy input density. If we continue to increase the input energy, the capillary flow will be less than the evaporation flow, the liquid level will retreat, and the evaporation interface will expand backward. The limit power density of sand, loam and clay was computed with some typical values. Given $r_1 = 40$ mm, $r_2 = 10$ m, the peak power density is: 811.4 kW m$^{-2}$, 77.9 kW m$^{-2}$, 12.2 kW m$^{-2}$ for sand, loam and clay respectively. Because although the maximum capillary force increases as particle size decreases, the permeability of soils decreases with particle size increasing. According to the analyses, for soils smaller particles, larger tube should be designed in order to achieve the same water harvesting rate.

In our laboratory experiments, the input power was only 15.9 kW m$^{-2}$, far below the maximum power density of the sand soil. The theoretical maximum mass flow rate under one sun is estimated to be 1.34 kg m$^{-2}$ h$^{-1}$ under one sun. We achieved 360 g h$^{-1}$ because of the limited size of the container and the mass loss during condensation, which indicate that there is still room for the improvement of our devices.

## 4 Conclusion

In summary, we present a novel method which can harvest water from the soils with low cost, simple structure and solar-driven devices. Our proposal may be suitable to provide cheap fresh water in the impoverished, arid, decentralized systems, and in post-disaster times. Both the model and the experiments predicted that the water could migrate to the vaporized area, resulted in continuously water harvesting. In the laboratory, we observed that this process is efficiently with respect to the sand soil under different water content and input power density. The maximum water mass

harvesting rate was 0.1 kg h$^{-1}$. In about 12 h, the total harvesting water could be as high as about 0.9

kg. Then the maximum water harvesting rata of the device under one sun energy flux (1 kW m$^{-2}$) is estimated to be about 0.36 kg h$^{-1}$ with a 1 m$^2$ solar concentrator. Our current experimental setup is possible to harvest 4.32 kg water for 12 hours under the above condition, which can meet the fresh water requirements of two people for one day.

**Acknowledgements**

This work was supported by the National Natural Science Foundation of China (51702357,

11602285) and the

Research Grant of China Aerospace Science and Technology Corporation (Y-KC-JT-00-013).

**Competing interests**

The authors declare no competing interests.

**Appendix A. Supplementary material**

Supplementary data to this article can be found online at

# Water harvesting from Soils by Solar-to-Heat Induced Evaporation and Capillary Water Migration

Xiaotian Li[†], Guang Zhang[†], Chao Wang, Lichen He, Yantong Xu, Rong Ma and Wei Yao*

*Department of Space sciences, Qian Xuesen Laboratory of Space Technology, China Academy of Space Technology, Beijing 100094, China.*

[†]*These authors contributed equally to this work.*

*Correspondence and requests for materials should be addressed to W. Y (yaowei@qxslab.cn)

## 1. Fabrication, characterization, and properties of the CNT films

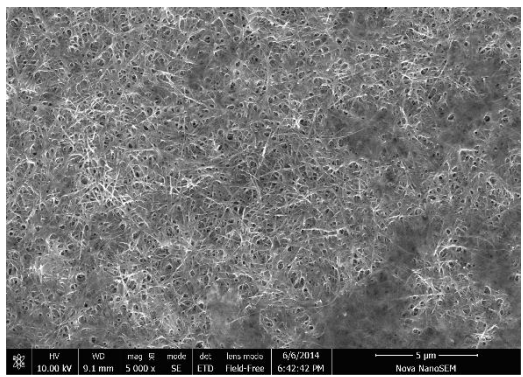

**Fig. S1.** Scanning electron microscope image of the CNT films.

CNT films were smeared on the inner face of the drill to increase the adsorption of the infrared radiation, thus improved the efficiency of converting optical energy to heat. The CNT was synthesized by the vapor chemical deposition method. [1,2]Then the as-prepared CNT samples were dispersed in ethanol by ultrasonic dispersion for five hours. Then a vacuum spray gun was used to spray the CNT films on the inner face of the drill. The scanning electron microscope image of the CNT films is shown in **Fig. S1**, which indicates that the drill is completely covered by the CNT films. The diameters and lengths of individual CNTs were about 10-20 nm and 100 to 900 μm, respectively. This method was reasonable because the infrared adsorption ratio of CNTs coating thinner than 2000 nm was found to be 95%[3]. In addition, the robust layers could withstand temperature up to 1000 ºC, which ensured its working performance during the experiment.

## 2. Photograph of the hollow tube

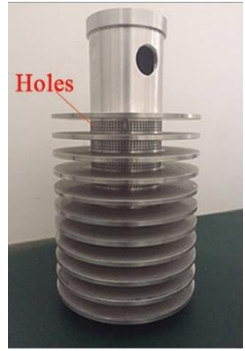

**Fig. S2.** Photograph of the hollow tube

## 3. Detailed Numerical simulation process by the COMSOL Multiphysics software

The geometry and boundary conditions is depicted in **Fig.S3**

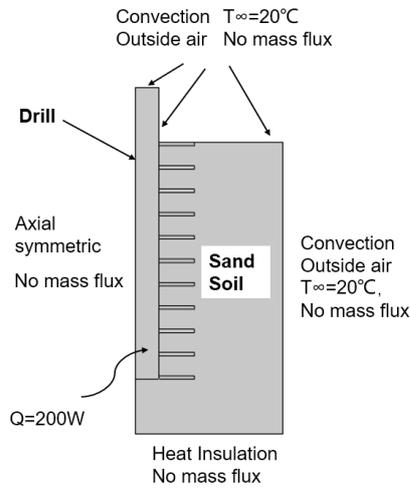

**Fig. S3** Geometry and boundary conditions

Mass equation can be described as follows:

$$\frac{\partial \theta_L}{\partial t} + \nabla(D_L \nabla \theta_L) = -\dot{m}$$

Where $\theta_L$ is the volume fraction of water, $D_L$ is the capillary diffusion coefficient, $\dot{m}$ is the evaporation rate, and since $\theta_L$ is a dimensionless number, the unit of $(\partial \theta_L)/\partial t$ is 1 /s, in order to ensure that the physical quantity $\theta_L$ is conserved, the other two units must be unified.

Therefore, after dimensional analysis, the unit of $D_L$ is m²/s, and the unit of $\dot{m}$ is 1/s.

Evaporation rate is described as follows:

$$\dot{m} = \begin{cases} \dfrac{k_{evap}\rho_L(\text{p}^* - \text{p}_G)}{\text{p}_G}, & \text{p}^* > \text{p}_G \\ 0, & \text{p}^* \leq \text{p}_G \end{cases}$$

Where $k_{evap}$ is the evaporation coefficient and is 10⁻⁶ 1/s. $\rho_L$ is the density of water,

$\text{p}_G$ is the vapor pressure. $\text{p}^*$ is the saturated vapor pressure, which is obtained from the

Antoine equation:

$$\log \text{p}^* = A - B/(T + C)$$

Where A, B, C are the Antoine constants and T is the gas temperature.

Energy equation can be described as follows:

$$c_{p,eff}\rho_{eff}\frac{dT}{dt} = \nabla(\lambda_{eff}\nabla\text{T}) - \dot{m}\Delta H$$

Where $c_{p,eff}$ is the equivalent specific heat capacity, $c_{p,eff} = (\theta_L\rho_L c_{p,L} + \theta_S\rho_S c_{p,S} + \theta_G\rho_G c_{p,G})/\rho_{eff}$, $\rho_{eff}$ is the equivalent density, $\rho_{eff} = \theta_L\rho_L + \theta_S\rho_S + \theta_G\rho_G$, λ_eff is the equivalent thermal conductivity, $\lambda_{eff} = \lambda_{dry} + \frac{\theta_L}{1-\theta_S}(\lambda_{wet} - \lambda_{dry})$. ΔH = 2257.2 kJ/kg (approximately 100 °C value). $\theta_S$, $\theta_L$, $\theta_G$ are the volume fractions of solids, liquids and gases, respectively.

Initial conditions:

The initial temperature is 20 ºC, and the initial water contents of soils are 5%, 10%, 20%.

Carbon Nanotube Films Covered Microchannel-Surface for Passive Electronic

Cooling Devices. *ACS Appl. Mater. Interfaces* **8**, 31202–31211 (2016).